\documentstyle[epsf,rotate,twocolumn,aps]{revtex}

\begin{document}
\draft              

\twocolumn[\hsize\textwidth\columnwidth\hsize\csname 
@twocolumnfalse\endcsname

%\twocolumn[\hsize\textwidth\columnwidth\hsize\csname @twocolumnfalse\endcsname

\title{Quasiparticle lifetime behaviour\\ 
in a simplified self--consistent T--matrix treatment \\ 
of the attractive Hubbard model in 2D}
%of the attractive Hubbard model in 2D:  Electronic density of states}

\author{M. Letz\cite{email} and R. J. Gooding}
%\author{M. Letz\cite{email} and R. J. Gooding}

\address{Dept. of Physics, Queen's University, Kingston, ON Canada K7L 3N6}

\date{\today}
\maketitle

%\widetext
\begin{abstract}
The attractive Hubbard model on a 2--D square lattice is studied at
low electronic densities using the ladder approximation for the
pair susceptibility. This model includes (i) the short coherence
lengths known to exist experimentally in the cuprate superconductors,
and (ii) two--particle bound states that correspond to electron pairs.  
We study the quasiparticle lifetimes in both non self--consistent and 
self--consistent theories, the latter including interactions between the 
pairs. We find that if we include the interactions between pairs the 
quasiparticle lifetimes vary approximately linearly with the inverse 
temperature, consistent with experiment.
\vskip 0.1 truecm
{\small Keywords:  attractive Hubbard model; quasiparticle properties}
\end{abstract}
\pacs{}
]
\vskip -0.5cm
%\twocolumn
\narrowtext

Numerous recent experiments ({\em e.g.}, neutrons, ARPES, optical,
NMR) have shown that in the high T$_c$ cuprate superconductors 
a so--called pseudogap is present \cite {houston}. This has led
to proposals that electron pairs form at temperatures well above the 
superconducting transition temperature. 
However, possibly due to phase fluctuations, a
macroscopic phase coherent wave function is not formed, so superconductivity
is not encountered \cite {emery}. 

This physics motivates our study of the attractive Hubbard model. The
Hamiltonian for this system is
\begin{equation}
H =-t\sum_{\langle ij \rangle, \sigma} 
( c^\dagger_{i,\sigma} c_{j,\sigma} + {\rm {h.c.}} )
~-~\mid U \mid~\sum_i n_{i,\uparrow} n_{i,\downarrow}
\label{eq:NUHM}
\end{equation}
where the lattice sites of a 2--D square lattice are labeled by $\{i\}$,
the lattice fermion operators are denoted by $c_{i,\sigma}$,
and neighbouring sites are represented by $\langle ij \rangle$.
For any nonzero $\mid U \mid$, two--particle bound states
appear, and are physically related to a pair of electrons lowering
the system's energy when they exist on the same lattice site.
This is certainly the simplest example of a model Hamiltonian which allows 
for one to study the interactions between such electron pairs.
Further, using a Gorkov derivation of the Ginzburg--Landau equations 
for an s--wave superconductor, one can show that for the values of 
$\mid U \mid / W$ that we are considering, 
$W = 8t$ being the (noninteracting) bandwidth, this model also reproduces 
the short coherence lengths of the Cooper pairs found in
the high $T_c$ cuprate superconductors \cite {feder}.

We employ the Brueckner Hartree--Fock theory to solve the
Bethe--Salpeter equation for the 2D attractive Hubbard model
in the ladder approximation \cite {FW},
reliable for this model at low electron (or hole) densities. 
This affords us the opportunity to investigate the influence of 
preformed pairs {\em with arbitrary lifetimes} on the normal state properties. 
Haussmann has argued \cite {haussmann93}
that when one solves this model in this approximation 
{\em self consistently}, one will include pair--pair interactions.
These interactions will be, in first order, of a repulsive nature 
between different pairs.

The equations for the Green's function ($G$), pair susceptibility ($\chi$), 
four--leg vertex function ($\Gamma$), and self energy ($\Sigma$), in a 
conserving approximation are well known:
\begin{eqnarray}
\label{eq:green}
G({\bf k},i \omega _n) &=& \left (G_0({\bf k},i \omega _n)^{-1} - \Sigma({\bf k},i
\omega _n) \right )^{-1} \\
\label{eq:chi}
\chi({\bf K},i \Omega _n) &=& \frac{-1}{N \beta} \sum_{m,{\bf k}}  
G({\bf K}-{\bf k},i \Omega _n - i \omega _m) G({\bf k},i \omega _m) \\
\label{eq:gamma}
\Gamma ({\bf K},i \Omega _n) &=& \mid U \mid / 
\left ( 1 + \mid U \mid \; \chi({\bf K},i \Omega _n) \right ) \\
\label{eq:self}
\Sigma({\bf k},i \omega _n) &=&  \frac{1}{N \beta} \sum_{m,{\bf q}}
\Gamma ({\bf k}+{\bf q },i \omega _n + i \omega _n) G({\bf q},i \omega _m)
\end{eqnarray}
where the wave vectors and Matsubara frequencies have their usual meaning.
In a non self--consistent theory, one replaces the full Green's functions $G$
in Eqs.~(3,5) with the noninteracting Green's functions $G_0$.

For full self consistency, this set of equations has to be solved iteratively.
Since such solutions are very difficult to obtain, we have investigated a 
simple approximation that allows for extensive numerical investigations of 
the resulting equations. To be specific, during the first step of the 
iteration process leading to self consistency,  we make an approximation for 
the pair susceptibility as being equal to the ${\bf k}$--average (denoted
from now on as overlined quantities, {\em e.g.} $\overline{\Gamma}$) of
the noninteracting pair susceptibility. We only calculate
the ${\bf k}$--averaged pair susceptibility during subsequent iterations
to self consistency. This leads to the following expressions
\begin{eqnarray}
&\overline{\Gamma}&(i \Omega _n)=\frac{1}{N}\sum_{\bf K}\Gamma
({\bf K},i \Omega _n)\approx \mid U \mid /(1+\mid U \mid\overline{\chi}
(i\Omega_n) ) \\
&{G}&({\bf k},i \omega _n)  \approx \left ({G_0}({\bf k},i \omega
_n)^{-1} - 
\overline{\Sigma}(i
\omega _n) \right )^{-1} 
\end{eqnarray}
which becomes very accurate in any of the following limits: large
$\mid U \mid / t$; high temperatures; large spatial dimensions. 
Implicit in this 
formulation is the assumption that the self energy is largely ${\bf k}$ 
independent.  We will discuss more fully the regimes of validity of this 
approximation in a future publication \cite {kaveragelang}. 

Representative results from our work are shown in two figures below.
In the first, we show the energy dependence of the imaginary part 
of the self energy. The non self--consistent curve shows the seemingly
unphysical result that the lifetime of the quasiparticles is
actually {\em shortest} at the fermi energy. This result may
be understood, at least in part, in terms the non self--consistent, 
non conserving theory of Ref. \cite {svr} --- the fermion density is 
suppressed as one approaches the Thouless criterion, and all
quasiparticles form two--particle bound states (this behaviour
survives when one treats the non self--consistent theory in
a conserving approximation \cite {lgm}). Our simplified self--consistent, 
conserving theory shows that this physics is completely
lost when interactions between the pairs are included --- now
one finds, similar to a usual fermi liquid, that the lifetimes of
quasiparticles are {\em longest} at the fermi energy. Quite simply,
this result reflects the absence of long--lived two--particle
bound states in a self--consistent theory.

Perhaps the most interesting result which follows from this work
involves the temperature dependence of the lifetime of quasiparticles.
As shown in Fig. 2, we find that the inverse of this lifetime
has a linear variation with temperature, similar
to the experimentally observed inelastic scattering rate
of the anomalous normal state of the high T$_c$ superconductors \cite {linres}.
The relationship of this result to different phenomenologies
for the anomalous normal state will be presented elsewhere 
\cite {kaveragelang}.

\begin{figure}
\unitlength1cm
\epsfxsize=13cm
\begin{picture}(7,7.5)
\put(-3.0,-1.5){\rotate[r]{\epsffile{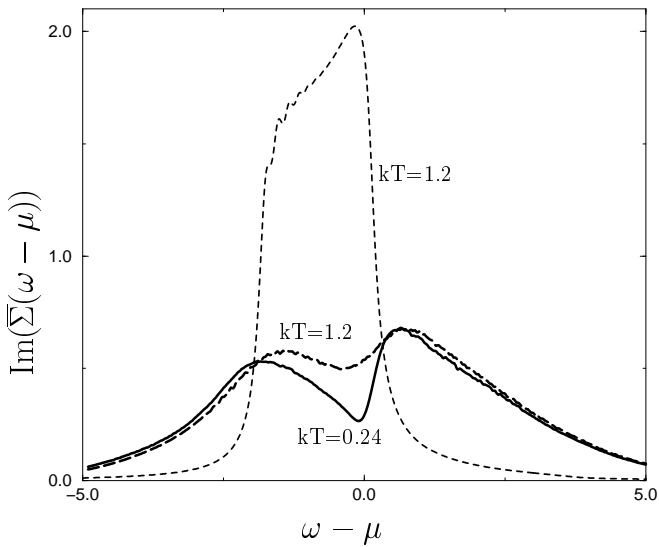}}}
\end{picture}
\caption{The imaginary part of the self energy
$\overline{\Sigma}(\omega - \mu)$ is plotted for two different temperatures,
$kT$ = 1.2 (dashed line) and $kT$ = 0.24 (solid line)
as a function of the frequency $\omega$.
For comparison we have 
plotted both the imaginary part of the non self--consistent self energy
for $kT$ = 1.2 (thin dashed line).  
All energies are in units of the transfer $t$ with the model parameters
given by $\mid U \mid / t = 8$, with electron filling $n = 0.2$.
\label{figa}
}
\end{figure}

\begin{figure}
\unitlength1cm
\epsfxsize=13cm
\begin{picture}(7,7.5)
\put(-3.0,-1.5){\rotate[r]{\epsffile{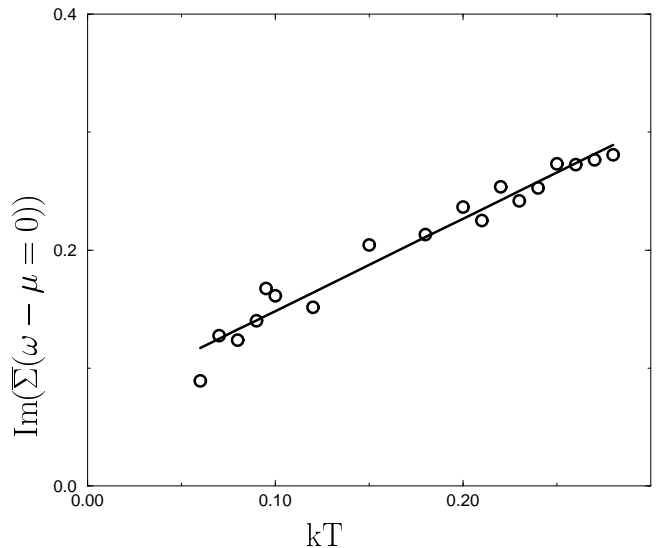}}}
\end{picture}
\caption{The imaginary part of the self energy at the
chemical potential, related to the inverse quasiparticle lifetime,
is plotted as a function of temperature. A linear variation 
seems to interpolate through the numerical data quite well (solid line).
\label{fig2}
}
\end{figure}

In conclusion, we have demonstrated the possibility of solving the
attractive Hubbard model in 2--D in the ladder approximation self consistently
by performing a k--average. Using this technique we include pair--pair
interactions which hinder the formation of infinite lifetime bound
states, and this leads to a quasiparticle lifetime that depends
inversely on temperature, similar to many experiments.

The authors wish to thank Frank Marsiglio and David Feder for numerous helpful
discussions. This work was supported by the NSERC of Canada.
M.L. acknowledges financial support from the DFG (Deutsche
Forschungsgemeinschaft).

%\end{thebibliography}

\vskip 2cm
%\newpage
%***********************************************************************
%{\Large \bf Figure captions} \\ \vskip 0.1cm
%***********************************************************************
%\begin{figure}
%\caption{
%}
%\label{fig1}
%\end{figure}

\end{document}